\documentclass[a4paper,11pt]{article} 

\usepackage[english]{babel}
\usepackage[a4paper]{geometry}
\usepackage{enumerate}  
\usepackage{amsmath}
\usepackage{pifont}
\usepackage{dsfont}
\usepackage{graphicx}
\usepackage{amsthm}
\usepackage{amssymb}

\usepackage{hyperref}    

             \let\l=\lambda
                          \let\r=\rho

  \def\v0{{\vec 0}}

\def\bal{{\bar \l}}

\def\indic{\hbox{\raise-2pt \hbox{\indbf 1}}}

\let\==\equiv

\let\0=\noindent

\def\*{{\hfill\break\null\hfill\break}}
\def\bra#1{{\langle#1|}}
\def\ket#1{{|#1\rangle}}

\def\tende#1{\,\vtop{\ialign{##\crcr\rightarrowfill\crcr
			\noalign{\kern-1pt\nointerlineskip}
			\hskip3.pt${\scriptstyle #1}$\hskip3.pt\crcr}}\,}
\def\otto{\,{\kern-1.truept\leftarrow\kern-5.truept\to\kern-1.truept}\,}

\def\sqt[#1]#2{\root #1\of {#2}}

\def\be{\begin{equation}}
	\def\ee{\end{equation}}
\def\bea{\begin{eqnarray}}\def\eea{\end{eqnarray}}
\def\bean{\begin{eqnarray*}}\def\eean{\end{eqnarray*}}
\def\bfr{\begin{flushright}}\def\efr{\end{flushright}}
\def\bc{\begin{center}}\def\ec{\end{center}}
\def\bal{\begin{align}} 
	\def\eal{\end{align}}

\def\spl#1\spl{\[ \begin{split}#1\end{split} \]}

\def\bd{\begin{description}}\def\ed{\end{description}}
\def\bv{\begin{verbatim}}\def\ev{\end{verbatim}}

\def\Halmos{\hfill\vrule height10pt width4pt depth2pt \par\hbox to \hsize{}}

\newtheorem{theorem}{Theorem}[section]
\newtheorem{proposition}{Proposition}[section]
\newtheorem{lemma}[proposition]{Lemma} 
\theoremstyle{remark}

\numberwithin{equation}{section}
\date{\today}     

\begin{document}

\title{Trial states for Bose gases: singular scalings and non-integrable potentials}
\author{Alessandro Olgiati\footnote{Institute of Mathematics, University of Z\"urich. Winterthurerstrasse 190, 8057 Z\"urich} }
%
%
\maketitle

\abstract{We review two results in which trial states for bosonic Hamiltonians were discussed. The problem of finding a trial state for a system with a hard-core potential in the Gross-Pitaevskii regime was recently solved by proving a link with the problem of finding a trial state for a system with a more regular potential in a less singular scaling, one of the type $N^{-1+3\beta}V(N^\beta\cdot)$ with $\beta\in(0,1)$. For both of these models we present the main result for the upper bound to the ground state energy, and discuss the key steps in the proof.}

\section{Introduction and main results}

The aim of this work is the review of the results \cite{BocBreCenSch-20,BasCenOlgPasSch-22b} on the ground state energy of dilute Bose gases with singular potentials. These investigations have a long history in mathematical physics, starting from \cite{Dys-57,LieYng-98,LieSeiYng-00} in which the leading order of the ground state energy was first computed. Later works on the Gross-Pitaevskii (GP) regime~\cite{LieSei-06,NamRouSei-16,MicNamOlg-19} extended the above leading order results to more general settings (see also the reviews \cite{LieSeiSolYng-05,Rou-20}). A number of proofs, in various scaling settings, were also obtained~\cite{ErdSchYau-08,Sei-11,GreSei-13,LewNamSerSol-15,BocBreCenSch-19,BocBreCenSch-20,NamTri-21,BreSchSch-22b} for the next-to-leading contributions to the ground state energy and to excited eigenvalues. These eventually culminated in the mathematical proofs of the Lee-Huang-Yang formula~\cite{YauYin-09,FouSol20,FouSol-21,BasCenSch21}, a second order ground state energy expansion in the thermodynamic limit which was initially conjectured in the physics literature \cite{Bog-47,LeeHuaYan-57} using the tools of Bogoliubov theory.

Extensions of the above results on the Bogoliubov correction for the ground state energy to the case of hard-core potentials have proven to be particularly challenging, especially as far as upper bounds are concerned. In fact, the rigorous justification of the Lee-Huang-Yang formula as an upper bound is still an open problem in the case of hard-core interaction (the lower bound was proven in \cite{FouSol-21}). The energy upper bound in the Gross-Pitaevskii regime was in turn obtained in the work \cite{BasCenOlgPasSch-22b}, whose proof strategy we review here. A key part of the result is the reduction of the highly singular initial problem (non-integrable potential and GP regime) to an effective model with bounded interaction potential and with a less singular scaling than GP, of the type considered in \cite{BocBreCenSch-20}. It is therefore very instructive to review the proof of the energy upper bound in \cite{BocBreCenSch-20} first. It is also worth mentioning that a similar trial state to the one we will discuss for the hard-core interaction was also recently used to prove an upper bound for the ground state energy of a dilute Bose gas in two dimensions \cite{FouGirJunMorOli-22}.

We consider bosonic systems trapped in the box $\Lambda=[-1/2,1/2]^{\times 3}$ with periodic boundary conditions (meaning that distances will be intended as measured on the unit torus). The class of Hamiltonians we first consider depends on a parameter $\beta \in(0,1]$ and reads
\begin{equation} \label{eq:H_N_beta}
H_N^{(\beta)}:= \sum_{j=1}^N -\Delta_{x_j}+\frac{1}{N}\sum_{i<j}^N N^{3\beta} V(N^\beta(x_i-x_j)).
\end{equation}
Under suitable assumption on $V$ (weaker than the ones specified below) this is a self-adjoint operator on a suitable dense subspace of the Hilbert space $L^2_\mathrm{sym}(\Lambda^N)$. The latter is the subspace of $L^2(\Lambda^N)$ consisting of functions of $N$ three-dimensional variables which are symmetric under permutations of any pair of variables.

The explicit $N$-dependence in the interaction term depends on the parameter $\beta$, and it represents a regime of strong interaction occurring at short length scales (of the order $N^{-\beta}$). For $\beta=1$ this is the Gross-Pitaevskii regime, a model for a highly diluted gas. The interaction potential $V$ in \eqref{eq:H_N_beta} is assumed to be measurable, non-negative, spherically symmetric, and compactly supported, with the further integrability requirement $V\in L^3(\mathbb{R}^3)$. We consider the ground state energy $E_N^{(\beta)}$ of $H_N^{(\beta)}$ under these assumptions.

It turns out that the expansion for the ground state energy exhibits important differences between the case $\beta\in(0,1)$ and $\beta=1$. We consider the former case in the next result, and discuss the case $\beta=1$ in Theorem \ref{thm:main_GP} and \ref{thm:main_hc} below. We will use the notations
\begin{equation}
\Lambda^*:= 2 \pi \mathbb{Z}^3,\qquad \Lambda_+^*:= \Lambda^*\setminus\{0\}
\end{equation}
to denote, respectively, the lattice of momenta dual to the torus $\Lambda$ and its version after the zero momentum has been removed.

\begin{theorem}[\textbf{Upper bound for $\beta\in(0,1)$}] \label{thm:main_beta}\mbox{} \\
	Let $V\in L^3(\mathbb{R}^3)$ be non-negative, spherically symmetric, and compactly supported. Then, if $\beta\in(0,1)$,
	\begin{equation} \label{eq:ub_beta}
		\begin{split}
			E_N^{(\beta)} \le\;& 4\pi \mathfrak{a}_N^{(\beta)}(N-1)\\
			&-\frac{1}{2}\sum_{p\in \Lambda_+^*}\left[ p^2+\widehat V(0)-\sqrt{|p|^4+2p^2 \widehat{V}(0)}-\frac{\widehat{V}^2(0)}{2p^2}\right]+ CN^{-\alpha},
		\end{split}
	\end{equation}
	for some $C,\alpha>0$. Here
	\begin{equation} \label{eq:a_beta}
	\begin{split}
	8\pi \mathfrak{a}_N^{(\beta)}:=\;&\widehat{V}(0)-\frac{1}{2N}\sum_{p\in \Lambda_+^*} \frac{\widehat V^2(p/N^\beta)}{p^2}\\
	&+\sum_{k=2}^{m_\beta}\frac{(-1)^k}{(2N)^k}\sum_{p\in\Lambda_+^*}\frac{\widehat V(p/N^\beta)}{p^2}\\
	&\quad\times\sum_{q_1,\dots, q_{k-1}\in \Lambda_+^*}\frac{\widehat V((p-q_1)/N^\beta)}{q_1^2} \prod_{i=1}^{k-2}\frac{\widehat V((q_i-q_{i+1}/N^\beta))}{q_{i+1}^2}\widehat V(q_{k-1}/N^\beta),
	\end{split}
	\end{equation}
	and $m_\beta$ is the largest integer such that $m_\beta \le 1/(1-\beta)+\min \{1/2,\beta/(1-\beta)\}$.
\end{theorem}

This result (together with a matching lower bound, as well as expansions for the low-energy excitation spectrum) was proven in \cite{BocBreCenSch-20}. We will review the main steps of its proof in Section~\ref{sect:proof_beta}.

It is worth remarking that the sum over $p\in \Lambda^*_+$ appearing on the right hand side of \eqref{eq:ub_beta} converges, and it yields a $\mathcal{O}(1)$ contribution to the ground state energy. The sum that defines $\mathfrak{a}_N^{(\beta)}$, in turn, coincides with $N$ times a Born expansion for the scattering length of the potential $N^{-1+3\beta} V(N^\beta \cdot)$. The $k$-th term in the sum is of the order $N^{k(\beta-1)}$ and, depending on the value of $\beta\in (0,1)$, a finite number of those terms yield a much larger contribution to the energy than the $\mathcal{O}(1)$-term in \eqref{eq:ub_beta}. The truncation $m_\beta$ is chosen precisely so as to ensure that further contributions to the Born series can be safely included in the remainder term in \eqref{eq:ub_beta}.

A formula similar to \eqref{eq:ub_beta} was proven in the case $\beta=1$. Here however every term of the Born expansion analogous to \eqref{eq:a_beta} is of the same order, and therefore the series cannot be truncated at the expense of negligible contributions. The whole sum nonetheless converges, and it yields the $s$-wave scattering length $\mathfrak{a}$ of the unscaled potential $V$. This is defined in terms of the zero-energy scattering equation
\begin{equation} \label{eq:zero_energy}
(-\Delta+V/2)f=0
\end{equation}
with boundary condition $f(x)\to 1$ as $|x|\to\infty$. For $x$ outside of the support of $V$ one has
\begin{equation}
f(x)=1-\frac{\mathfrak{a}}{|x|}
\end{equation}
for some non-negative number $\mathfrak{a}$ which is called by definition the scattering length of $V$. Equivalently, $\mathfrak{a}$ can be defined as the number satisfying
\begin{equation}
8 \pi \mathfrak{a}=\int_{\mathbb{R}^3} V f.
\end{equation}
The following result states the upper bound for the ground state energy $E_N^{(1)}$ of $H_N^{(1)}$. It was proven first in \cite{BocBreCenSch-19}, together with a matching lower bound and with analogous expansions for the excitation spectrum.

\begin{theorem}[\textbf{Upper bound in the Gross-Pitaevskii regime, integrable case}] \label{thm:main_GP}\mbox{} \\
	Let $V \in L^3(\mathbb{R}^3)$ be non-negative, spherically symmetric, compactly supported, and with scattering length $\mathfrak{a}$. Then there exists $C>0$ such that
	\begin{equation} \label{eq:ub_GP}
	\begin{split}
	E_N^{(1)} \le\;& 4\pi \mathfrak{a}(N-1)+e_\Lambda \mathfrak{a}^2\\
	&-\frac{1}{2}\sum_{p\in \Lambda_+^*}\left[ p^2+8\pi\mathfrak{a}-\sqrt{|p|^4+ 16\pi \mathfrak{a}p^2}-\frac{(8\pi \mathfrak{a})^2}{2p^2}\right]+ CN^{-1/4},
	\end{split}
	\end{equation}
	having defined
	\begin{equation*}
	e_\Lambda:=2- \lim_{M\to\infty} \sum_{\substack{p\in\mathbb{Z}^3\setminus\{0\}\\|p_1|,|p_2|,|p_3|\le M}}	\frac{\cos(|p|)}{p^2},
	\end{equation*}
	where in particular the limit exists.
\end{theorem}

We will not review here the proof of Theorem \ref{thm:main_GP}, which the reader can find in \cite{BocBreCenSch-19}. It is based on a highly non-trivial modification of the proof of Theorem \ref{thm:main_beta}.

The result of Theorem \ref{thm:main_GP} (together with a matching lower bound) makes it manifest that, in this dilute regime, the only relevant interaction parameter is the scattering length. The constant $e_\Lambda$ appearing in \eqref{eq:ub_GP} is a finite size effect due to the fact that the system is trapped in a box of unit volume while the scattering length $\mathfrak{a}$ is defined through the solution to \eqref{eq:zero_energy} on the whole $\mathbb{R}^3$.

It is an interesting problem to ask whether, both in Theorem \ref{thm:main_beta} and \ref{thm:main_GP}, the assumption $V\in L^3(\mathbb{R}^3)$ can be relaxed. In fact, as far as the GP case is concerned, the scattering length, and thus all the relevant quantities in the right hand side of \eqref{eq:ub_GP}, are well-defined under much weaker assumptions, even including some non-locally integrable $V$. The most natural example of those is the so-called hard-core potential
\begin{equation} \label{eq:hc_potential}
V_\mathrm{HC}(x)= \begin{cases}
+\infty &\text{ if }|x|\le \mathfrak{a}\\
0 & \text{ if } |x| > \mathfrak{a},
\end{cases}
\end{equation}
for which the radius $\mathfrak{a}$ of the support coincides with the scattering length. The extension of Theorem \ref{thm:main_GP} to the hard-core case, which we present below, was recently obtained in \cite{BasCenOlgPasSch-22b}.

The rigorous way to define a model with an interaction of the type $V_\mathrm{HC}$ is to restrict the Hilbert space by means of a boundary condition, namely to study the Hamiltonian
\begin{equation}\label{eq:H_N_HC}
H_N^\mathrm{HC}:= \sum_{j=1}^N -\Delta_{x_j}
\end{equation}
with form domain
\begin{equation}  \label{eq:D_HC}
\begin{split}
\mathcal{D}[H_N^\mathrm{HC}]:=\;& \Big\{ \psi_N \in L^2_\mathrm{sym}(\Lambda^N)\;|\; \psi_N(x_1,\dots,x_N)=0 \text{ if } \exists\, i\ne j\\
&\qquad\qquad\qquad\qquad  \text{ s.t. }|x_i-x_j| \le \frac{\mathfrak{a}}{N} \Big\}\cap H^1(\Lambda^N).
\end{split}
\end{equation}
The fact that the hard-core radius appears divided by $N$ effectively sets the problem in the Gross-Pitaevskii regime. The ground state energy is then
\begin{equation}
E_N^\mathrm{HC}=\inf_{\psi_N\in\mathcal{D[H_N^\mathrm{HC}]}}\frac{\sum_{j=1}^N \int_{\Lambda^N}|\nabla_{x_j}\psi_N|^2}{\int_{\Lambda^N}|\psi_N|^2},
\end{equation}
for which we have the following upper bound.

\begin{theorem} [\textbf{Upper bound in the Gross-Pitaevskii regime, hard-core case}] \label{thm:main_hc}\mbox{} \\
	There exist $C,\varepsilon>0$ such that
	\begin{equation} \label{eq:ub_HC}
	\begin{split}
	E_N^{\mathrm{HC}} \le\;& 4\pi \mathfrak{a}(N-1)+e_\Lambda \mathfrak{a}^2\\
	&-\frac{1}{2}\sum_{p\in \Lambda_+^*}\left[ p^2+8\pi\mathfrak{a}-\sqrt{|p|^4+ 16\pi \mathfrak{a}p^2}-\frac{(8\pi \mathfrak{a})^2}{2p^2}\right]+ CN^{-\varepsilon},
	\end{split}
	\end{equation}
	where $\mathfrak{a}$ is the quantity appearing in the hard-core boundary condition \eqref{eq:D_HC}.
\end{theorem}

We will review in Section \ref{sect:proof_hc} the main steps of the proof of \eqref{eq:ub_HC} from \cite{BasCenOlgPasSch-22b}. An important part of the discussion will be the link between the problem of finding a trial state for $H_N^\mathrm{HC}$ and that of finding a trial state for (a suitable modification of) $H_N^{(\beta)}$.

\section{Proof of Theorem \ref{thm:main_beta}: trial state for $\beta\in(0,1)$} \label{sect:proof_beta}

We associate to each element $p\in\Lambda^*$ the eigenfunction $u_p=e^{ip\cdot x}$ of the momentum operator. The function $u_0=1$ has a special role compared to all other plane waves, since the ground state of $H_N^{(\beta)}$ exhibits Bose-Einstein condensation into $u_0$, while a subleading fraction of particles occupies the orbitals $\{u_p\}_{p\ne0}$. Rigorous proofs of BEC were obtained in \cite{BocBreCenSch-20} for $\beta\in(0,1)$ and in \cite{LieSei-02,NamRouSei-16,BocBreCenSch-18,NamNapRicTri-20,Hai-21,BreSchSch-22a} for $\beta=1$. These results contain important guiding principles for the upper bound we are reviewing, but we shall not review them in detail here.

\subsubsection*{The excitation Hamiltonian}

It is convenient to rewrite $H_N^{(\beta)}$ in the form
\begin{equation}
H_N^{(\beta)}=\sum_{p\in\Lambda^*} p^2 a^*_p a_p + \frac{1}{2N}\sum_{p,q,r \in \Lambda^*}\widehat V(r/N^\beta) a^*_p a^*_q a_{q-r}a_{p+r},
\end{equation}
having used the notation $a_p = a(u_p)$ and $a_p^*=a^*(u_p)$ for standard bosonic creation and annihilation operators. In order to extract from $H_N^{(\beta)}$ the energy of the Bose-Einstein condensate we will conjugate $H_N^{(\beta)}$ with the excitation map $U_N$ introduced in \cite{LewNamSerSol-15}. This is a unitary operator
\begin{equation}
U_N: L^2_\mathrm{sym}(\Lambda^N) \to \mathcal{F}^{\le N}_+,
\end{equation}
that maps to the truncated Fock space of excited modes
\begin{equation}
\mathcal{F}^{\le N}_+ = \mathbb{C}\oplus\bigoplus_{n=1}^N \left( L^2(\Lambda)\cap \{u_0\}^\perp\right)^{\otimes_\mathrm{sym}n}.
\end{equation}
The action of $U_N$ is given by definition by
\begin{equation} \label{eq:def_U_N}
U_N\psi_N= \bigoplus_{n=0}^N (1-\ket{u_0} \bra{u_0})^{\otimes n} \frac{a_0^{N-n}}{\sqrt{(N-n)!}}\psi_N,
\end{equation}
and it satisfies the relations
\begin{equation} \label{eq:action_U_N}
\begin{split}
U_N a_0^* a_0 U_N^*=\;&N-\mathcal{N}_+\\
U_N a^*_p a_0 U_N^*=\;&a^*_p \sqrt{N-\mathcal{N}_+}=:\sqrt{N}b^*_p\\
U_N a^*_0 a_p U_N^*=\;&\sqrt{N-\mathcal{N}_+}a_p =:\sqrt{N}b_p\\
U_N a^*_p a_q U_N^*=\;&a^*_p a_q
\end{split}
\end{equation}
for all $p,q \in \Lambda^*_+$, having introduced the notation
$$\mathcal{N}_+= \sum_{p\in\Lambda^*_+} a^*_p a_p$$
for the number operator for excited modes.
The relations in \eqref{eq:action_U_N} define modified creation and annihilation operators $b_p,b^*_p$, that is, creation and annihilation operators that preserve the space $\mathcal{F}_+^{\le N}$ and coincide to good approximation with, respectively, $a_p,a^*_p$ on states with few excitations. It is then a straightforward computation to deduce that the excitation Hamiltonian
\begin{equation}
\mathcal{L}_N^{(\beta)}:= U_N H_N^{(\beta)} U_N^*
\end{equation}
reads
\begin{equation}
\mathcal{L}_N^{(\beta)}= \mathcal{L}_N^{0,(\beta)}+\mathcal{L}_N^{2,(\beta)}+\mathcal{L}_N^{3,(\beta)}+\mathcal{L}_N^{4,(\beta)}
\end{equation}
with
\begin{equation}
\begin{split}
\mathcal{L}_{N}^{0,(\beta)}=\;&\frac{N-1}{2N}\widehat V(0)(N-\mathcal{N}_+)+\frac{\widehat V(0)}{2N}\mathcal{N}_+ (N-\mathcal{N}_+)\\
\mathcal{L}_N^{2,(\beta)}=\;&\sum_{p \in \Lambda^*_+}p^2 a^*_p a_p+ \sum_{p\in\Lambda_+^*} \widehat V(p/N^{\beta})\left[ b^*_p b_p-\frac{1}{N}a^*_p a_p \right]\\
&+\frac{1}{2}\sum_{p\in \Lambda_+^*}\widehat V(p/N^\beta)\left[  b^*_p b^*_{-p}+b_p b_{-p}\right]\\
\mathcal{L}_N^{3,(\beta)}=\;&\frac{1}{\sqrt{N}}\sum_{p,q \in \Lambda_+^*,\; p+q \ne 0} \widehat V(p/N^\beta)\left[ b^*_{p+q}a^*_{-p}a_q +a^*_{q}a_{-p}b_{p+q} \right]\\
\mathcal{L}_N^{4,(\beta)}=\;&\frac{1}{2N}\sum_{\substack{p,q \in \Lambda_+^*,\; r\in \Lambda^*\\r \ne -p,-q}}\widehat V(r/N^\beta) a^*_{p+r}a^*_q a_p a_{q+r}.
\end{split}
\end{equation}

Computing the expectation of $\mathcal{L}^{(\beta)}_N$ on the vacuum vector $\Omega_+ \in \mathcal{F}_+^{\le N}$ gives the upper bound
\begin{equation} \label{eq:first_ub}
E^{(\beta)}_N \le \frac{N-1}{2} \widehat V(0).
\end{equation}
This reproduces the energy in \eqref{eq:H_N_beta} to leading order only, clearly indicating that we are neglecting contributions from the parts of $\mathcal{L}_N^{(\beta)}$ that vanish on the vacuum. We will show below that the first step to extract such contributions is to implement a short-scale correlation structure modeled by the solution to the scattering equation \eqref{eq:zero_energy}.

\subsubsection*{First quadratic correlation: short length scale structure}

We consider a slight modification of the zero-energy scattering equation \eqref{eq:zero_energy}, namely, we assume that $f_l^{(\beta)}$ is the ground state of the Neumann problem
\begin{equation} \label{eq:f_l}
\begin{cases}
\left( -\Delta+\frac{N^{3\beta-1}}{2}V(N^\beta x) \right) f^{(\beta)}_{l}(x)=\lambda_l^{(\beta)} f_l^{(\beta)}(x)\quad &|x|\le l\\
\partial_{|x|} f_l^{(\beta)}(x)=0\quad& |x|=l\\
f_l^{(\beta)}(x)=1\quad & |x|\ge l
\end{cases}
\end{equation}
for some radius $l>0$. It turns out (see \cite[Lemma 3.1]{BocBreCenSch-20}) that $f_l^{(\beta)}$ and $\lambda_l^{(\beta)}$ satisfy the properties presented in the following result.

\begin{lemma}
	Let $V$ satisfy the assumptions of Theorem \ref{thm:main_beta} and $0<l<1/2$. Then $0\le f_l^{(\beta)}\le 1$ and
	\begin{equation}
	\lambda_l^{(\beta)}=\frac{3\widehat V(0)}{8\pi Nl^3}(1+\mathcal{O}(N^{\beta-1})).
	\end{equation}
	Moreover, defining the function $w_l^{(\beta)}=1-f_l^{(\beta)}$, there exists $C>0$ such that 
	\begin{equation} \label{eq:decay_w_beta}
	w_l^{(\beta)}(x)\le \frac{C}{N(|x|+N^{-\beta})}
	\end{equation}
	for all $|x|\le l$ and
	\begin{equation} 
	|\widehat{w}^{(\beta)}(p)|\le \frac{C}{Np^2}
	\end{equation}
	for all $p\in\Lambda_+^*$.
\end{lemma}

Let us then define the Fourier coefficients
\begin{equation*}
\eta_p=-N \widehat{w_l^{(\beta)}}(p).
\end{equation*}
The scattering equation in \eqref{eq:f_l} is expressed in Fourier space in terms of $\eta_p$ as
\begin{equation} \label{eq:scat_eta}
\begin{split}
p^2 \eta_p+\frac{1}{2}&\widehat V(p/N^\beta)+\frac{1}{2N}\sum_{q \in \Lambda^*} \widehat V((p-q)/N^\beta)\eta_q \\
=\;& N \lambda_l^{(\beta)}\widehat{\chi_l}(p)+\lambda_{l}^{(\beta)}\sum_{q\in \Lambda^*}\widehat{\chi_l}(p-q)\eta_q,
\end{split}
\end{equation}
where $\chi_l$ is the characteristic function of the ball of radius $l$. We further define the operator
\begin{equation*}
B_\eta=\frac{1}{2}\sum_{p\in \Lambda_+^*} \eta_p \left( b^*_p b^*_{-p} - b_p b_{-p}\right)
\end{equation*}
and the unitary operator $e^{B_\eta}$ on $\mathcal{F}_+^{\le N}$.

Conjugating $\mathcal{L}_N^{(\beta)}$ with $e^{B_\eta}$, i.e. considering the operator
\begin{equation} \label{eq:G_N^beta}
\mathcal{G}^{(\beta)}_N = e^{-B_\eta} \mathcal{L}_N^{(\beta)} e^{B_\eta} = e^{-B_\eta} U_N H_N^{(\beta)}U_N^* e^{B_\eta},
\end{equation}
implements a two-body correlation structure modeled by $w_l^{(\beta)}$, and it allows to extract important contributions to the ground state energy. Since, by \eqref{eq:decay_w_beta}, the function $w_l^{(\beta)}$ decays on distances of order $N^{-\beta}$, it has at least heuristically the right behavior in order to correct the singular scaling created by the potential $N^{-1+3\beta}V(N^\beta\cdot)$. The precise statement is given by the next Proposition. We will use the notations $\sigma_p=\sinh(\eta_p)$ and $\gamma_p=\cosh(\eta_p)$, as well as
\begin{equation} \label{eq:N_+_K}
 \mathcal{K}= \sum_{p \in \Lambda_+^*} p^2 a^*_p a_p.
\end{equation}

\begin{proposition} \label{prop:quadratica_beta}
Under the same assumptions of Theorem \ref{thm:main_beta}, and if $l>0$ is small enough, we have
\begin{equation} \label{eq:expansion_G_beta}
\mathcal{G}_N^{(\beta)}= C_{N}^{(\beta)}+\mathcal{Q}_N^{(\beta)}+\mathcal{E}_N^{(\beta)},
\end{equation}
where
\begin{equation} \label{eq:C_N^beta}
\begin{split}
C_N^{(\beta)}=\;&\frac{(N-1)}{2}\widehat V(0)+\sum_{p\in\Lambda_+^*}\bigg[p^2 \sigma_p^2+\widehat V(p/N^\beta)(\sigma_p^2+\sigma_p\gamma_p)\\
&\qquad\qquad\qquad \qquad \quad+\frac{1}{2N}\sum_{q\in\Lambda_+^*}\widehat V ((p-q)/N^\beta)\eta_p\eta_q \bigg],\\
\mathcal{Q}_N^{(\beta)}=\;&\sum_{p\in\Lambda_+^*}\left[F_pa^*_p a_p+\frac{1}{2} G_p(b^*_{p} b^*_{-p}+b_pb_{-p})\right]
\end{split}
\end{equation}
with
\begin{equation} \label{eq:F_p_G_p}
\begin{split}
F_p=\;&p^2(\sigma_p^2+\gamma_p^2)+\widehat V(p/N^\beta)(\sigma_p+\gamma_p)^2\\
G_p=\;&2p^2 \sigma_p\gamma_p+\widehat V(p/N^\beta)(\sigma_p+\gamma_p)^2+\frac{1}{N}\sum_{q\in \Lambda_+^*}\widehat V((p-q)/N^\beta)\eta_q,
\end{split}
\end{equation}
and
\begin{equation}
\pm \mathcal{E}_N^{(\beta)}\le \frac{C}{N^{(1-\beta)/2}}(\mathcal{K}+1)(\mathcal{N}_++1).
\end{equation}
\end{proposition}

The proof of Proposition \ref{prop:quadratica_beta} is based on a careful analysis of the action of $e^{B_\eta}$ on all the terms of $\mathcal{L}_N^{(\beta)}$. The action of $e^{B_\eta}$ is in fact approximately that of a Bogoliubov transformation, up to reminders that emerge because the $b,b^*$-operators preserve the truncated space $\mathcal{F}_+^{\le N}$. All the terms appearing in $C_N^{(\beta)}$ and $\mathcal{Q}_N^{(\beta)}$ are in fact analogous to those which would be obtained by formally replacing the $b,b^*$'s by $a,a^*$'s, and by using the exact expressions for the action of a Bogoliubov transformation. The fact that that the $b,b^*$'s are only approximately a CCR algebra requires to control a larger number of remainder terms.  The reader can find the detailed proof of Proposition \ref{prop:quadratica_beta} in \cite[Section 7]{BocBreCenSch-20} for the case of $V$ of the type $\kappa \widetilde{V}$ with $\kappa$ small enough. The modifications needed in order to remove such an assumption where later introduced in \cite{BocBreCenSch-19} for the case $\beta=1$, and then discussed in \cite[Section 5]{BasCenOlgPasSch-22b} for the case of $\beta\in(0,1)$.

It is also worth remarking that, in this $\beta\in(0,1)$ regime, the operator $e^{-B_\eta} \mathcal{L}_N^{3,(\beta)}e^{B_\eta}$ is included in the remainder $\mathcal{E}_N^{(\beta)}$. In the case $\beta=1$, in turn, further cubic (and quartic) operators appear in the right hand side of \eqref{eq:expansion_G_beta}. This considerably complicates the subsequent steps.

One can now compute the expectation of $\mathcal{G}_N^{(\beta)}$ on $\Omega_+\in\mathcal{F}_+^{\le N}$, i.e. the expectation of $H_N^{(\beta)}$ on the trial state $U_N^* e^{B_\eta} \Omega_+$. The contribution of $\mathcal{Q}_N^{(\beta)}$ then vanishes, and some algebraic manipulations on the constant $C_N^{(\beta)}$ give the upper bound
\begin{equation}
 E_N^{(\beta)} \le 4\pi (N-1) a_N^{(\beta)} +C_l,
\end{equation}
where $a_N^{(\beta)}$ was defined in \eqref{eq:a_beta}, and $C_l$ is a constant depending on $l$ only. This captures the order of the correction to the energy in \eqref{eq:ub_beta}, but it stops short of reproducing the right constant. The effect of the correlations implemented through $B_\eta$ has been to replace with $a_N^{(\beta)}$ the first order Born approximation $\widehat V(0)$ appearing in the worse upper bound \eqref{eq:first_ub}. The missing contributions in order to recover the sum in the right hand side of \eqref{eq:ub_beta} come from the quadratic Hamiltonian $\mathcal{Q}_N^{(\beta)}$.

\subsubsection*{Diagonalization of $\mathcal{Q}_N^{(\beta)}$}

We last conjugate $\mathcal{G}_N^{(\beta)}$, and in particular $\mathcal{Q}_N^{(\beta)}$, with another quadratic transformation. This will have the property of bringing $\mathcal{Q}_N^{(\beta)}$ to the diagonalized form $E_0+\sum_p \omega_p a^*_p a_p$ for some $E_0\in\mathbb{R}$  and $ \omega_p\ge0$. The minimization will then be straightforwardly performed by computing the expectation on the vacuum.

Given $F_p$ and $G_p$ from \eqref{eq:F_p_G_p} we define
\begin{equation}
\tau_p=\frac{1}{4}\log \frac{1-G_p/F_p}{1+G_p/F_p}.
\end{equation}
Using the precise form of $F_p$ and $G_p$ and the scattering equation \eqref{eq:scat_eta} one sees (see \cite[Section 5]{BocBreCenSch-20}) that the logarithm is well defined, and
moreover
\begin{equation}
|\tau_p|\le \frac{C}{|p|^4}.
\end{equation}                                                                             
This shows that $\tau_p$ has a much better decay behavior than $\eta_p$ at momenta of the order $N^{\beta}$ (the scale of momenta at which the interaction occurs), and it thus models correlations on a much larger length scale than $N^{-\beta}$. 
We then define
\begin{equation}
B_\tau=\frac{1}{2}\sum_{p \in \Lambda_+^*}\tau_p (b^*_p b^*_{-p}-b_p b_{-p})
\end{equation}
and we consider the unitary operator $e^{B_\tau}$ on $\mathcal{F}_+^{\le N}$.

\begin{proposition} \label{prop:quadratica_tau}
	Let $\mathcal{G}_N^{(\beta)}$ be defined in \eqref{eq:G_N^beta} and ${C}_N^{(\beta)},\mathcal{Q}_N^{(\beta)}$ in \eqref{eq:C_N^beta}. Under the same assumptions of Theorem \ref{thm:main_beta} we have
	\begin{equation} \label{eq:diagonalization}
	e^{-B_\tau} \mathcal{Q}_N^{(\beta)}e^{B_\tau}=\frac{1}{2}\sum_{p\in\Lambda^*_+}\left[ -F_p+\sqrt{F_p^2-G_p^2} \right]+ \sum_{p \in \Lambda_+^*}\sqrt{F_p^2-G_p^2}a^*_p a_p+\delta_{N}^{(\beta)}
	\end{equation}
	with
	\begin{equation}
	\pm \delta_{N}^{(\beta)} \le \frac{C}{N}(\mathcal{N}_++1)(\mathcal{K}+1).
	\end{equation}
	Moreover, for $C_N^{(\beta)}$ defined in \eqref{eq:C_N^beta} and $\mathfrak{a}_N^{(\beta)}$ defined in \eqref{eq:a_beta}, we have
	\begin{equation} \label{eq:constants}
	\begin{split}
	C_N^{(\beta)}&+\frac{1}{2} \sum_{p \in \Lambda_+^*}\left[ -F_p+\sqrt{F_p^2-G_p^2} \right]\\
	=\;&4\pi (N-1)\mathfrak{a}_N^{(\beta)}-\frac{1}{2}\sum_{p\in \Lambda_+^*}\left[ p^2+\widehat V(0)-\sqrt{|p|^4+2p^2 \widehat{V}(0)}-\frac{\widehat{V}^2(0)}{2p^2}\right]\\
	&+\mathcal{O}(N^{-\alpha})
	\end{split}
	\end{equation}
	for some $\alpha>0$. Finally,
	\begin{equation} \label{eq:final}
	\begin{split}
	e^{-B_\tau} \mathcal{G}_{N}^{(\beta)}e^{B_\tau}=\;&4\pi (N-1)\mathfrak{a}_N^{(\beta)}\\
	&-\frac{1}{2}\sum_{p\in \Lambda_+^*}\left[ p^2+\widehat V(0)-\sqrt{|p|^4+2p^2 \widehat{V}(0)}-\frac{\widehat{V}^2(0)}{2p^2}\right]\\
	&+\sum_{p \in \Lambda_+^*}\sqrt{F_p^2-G_p^2}a^*_p a_p+\theta_{N}^{(\beta)}
	\end{split}
	\end{equation}
	with
	\begin{equation}
	\pm \theta_{N}^{(\beta)} \le \frac{C}{N^{\alpha}}(\mathcal{N}_++1)(\mathcal{K}+1).
	\end{equation}
\end{proposition}

The proof of Proposition \ref{prop:quadratica_tau} can be found in \cite[Section 5]{BocBreCenSch-20}. In order to show \eqref{eq:diagonalization} it is once again very useful to argue that the operator $e^{B_\tau}$ acts approximately as a Bogoliubov transformation. Once \eqref{eq:diagonalization} is proven, the quantity in \eqref{eq:constants} is the only non-negligible contribution to the expectation value of $e^{-B_\tau}\mathcal{G}_N^{(\beta)}e^{B_\tau}$ on the vacuum. Eq. \eqref{eq:final} follows then by keeping track of the action of $e^{B_\tau}$ on the remainder term $\mathcal{E}_N^{(\beta)}$ in $\mathcal{G}_N^{(\beta)}$.

Computing the expectation of $e^{-B_\tau} \mathcal{G}_N^{(\beta)} e^{B_\tau}$ on $\Omega_+ \in \mathcal{F}_+^{\le N}$, i.e., the expectation of $H_N^{(\beta)}$ on $U_N^* e^{B_\eta} e^{B_\tau}\Omega_+$, completes the proof of Theorem \ref{thm:main_beta}.

\section{Proof of Theorem \ref{thm:main_hc}: trial state with hard-core potential} \label{sect:proof_hc}

Writing a trial state for $H_N^\mathrm{HC}$ requires a change of paradigm with respect to the case of integrable $V$. A trivial trial state for $H_N^{(\beta)}$ was the function $\psi_N=U_N^* \Omega_+ \in L^2_\mathrm{sym}(\Lambda^N)$ which is everywhere equal to 1. This does not belong to the domain \eqref{eq:D_HC}, and neither do the more and more refined trial states reviewed in Section \ref{sect:proof_beta}, obtained by acting on $\Omega_+$ with unitary operators implementing quadratic (or cubic, in the case of $\beta=1$~\cite{BocBreCenSch-19}) correlations. A stronger correlation structure needs to be imposed in order to satisfy the condition in \eqref{eq:D_HC}.

\subsubsection*{The (Mott-Dingle-)Jastrow function}

Let us define the (non-normalized) wave-function
\begin{equation} \label{eq:jastrow}
\psi_N^{(\mathrm{J})}(x_1,\dots,x_N)= \prod_{i<j} f_\ell(x_i-x_j),
\end{equation}
where $f_\ell$ is the ground state of the Neumann problem with hard-core boundary condition
\begin{equation} \label{eq:f_ell}
\begin{cases}
-\Delta f_\ell(x)=\lambda_\ell f_\ell(x)\quad &\mathfrak{a}/N \le |x| \le \ell\\
f_\ell(x)=0\quad &|x|\le \mathfrak{a}/N\\
\partial_{|x|} f_\ell(x)=0\quad &|x|=\ell\\
f_\ell(x)=1 \quad &|x|\ge \ell,
\end{cases}
\end{equation}
with eigenvalue $\lambda_\ell$. Functions of the type $\psi^{(\mathrm{J})}_N$, with different choices of the correlation factor $f_\ell$, were considered first by Mott~\cite{Mot-49}, Dingle~\cite{Din-49}, and Jastrow~\cite{Jas-55} as trial states for bosonic Hamiltonians. The first rigorous upper bound to the ground state energy of a dilute Bose gas was obtained by Dyson~\cite{Dys-57} using a slightly different version of \eqref{eq:jastrow}, one which was not symmetric and which included correlations between nearest neighbouring particles only.

The most important observation about $\psi^{(\mathrm{J})}_N$ is that it belongs to the hard-core domain \eqref{eq:D_HC}, due to the vanishing of $f_\ell(x)$ for $|x|\le\mathfrak{a}/N$. Further properties of $f_\ell$ and $\lambda_\ell$, presented in the next Lemma, allow to compute the energy of $\psi^{(\mathrm{J})}_N$.

\begin{lemma} \label{lemma:f_ell}
	If $\ell <1/2$ with $N\ell\to\infty$, we have $0\le f_\ell\le1$ and
	\begin{equation}
		\lambda_\ell=\frac{3\mathfrak{a}}{N\ell^3}\left[ 1+\frac{9}{5}\frac{\mathfrak{a}}{N \ell}+\mathcal{O}\left( \frac{\mathfrak{a}^2}{N^2\ell^2} \right) \right].
	\end{equation}
	Moreover, if $w_\ell=1-f_\ell$, we have
	\begin{equation}
		\begin{cases}
			w_\ell(x)=1&\quad |x|\le \frac{\mathfrak{a}}{N}\\
			w_\ell(x)\le\frac{C\mathfrak{a}}{N|x|}&\quad \frac{\mathfrak{a}}{N}\le|x|\le \ell\\
			w_\ell(x)=0&\quad |x|\ge \ell,
		\end{cases}
	\end{equation}
	and the same properties hold if $w_\ell$ is replaced by $u_\ell=1-f_\ell^2$.
\end{lemma}

The reader can find the proof of Lemma \ref{lemma:f_ell} in \cite[Appendix A]{BasCenOlgPasSch-22b}.

It is well-known that the expectation of $H_N^\mathrm{HC}$ on (the normalized version of) $\psi_N^{(\mathrm{J})}$ correctly reproduces the leading order of the ground state energy. The explicit calculations can be found, e.g., in~\cite{BasCenOlgPasSch-22a} in the case of the thermodynamic limit (see also \cite{MicNamOlg-19} for a similar calculation in the case of two bosonic species). The resulting upper bound is
\begin{equation} \label{eq:ub_solo_jastrow}
	E_N^\mathrm{HC}\le 4\pi \mathfrak{a} N +\frac{C}{\ell}+C N \ell^2.
\end{equation}
This indeed agrees with \eqref{eq:ub_HC} to the leading order, as long as $\mathfrak{a}/N \ll \ell\ll 1$. However, no choice of $\ell$ makes the remainder in \eqref{eq:ub_solo_jastrow} of order $\mathcal{O}(1)$, which would be the order of the correction to \eqref{eq:ub_HC}. This is due to a trade-off on how to choose $\ell$:
\begin{itemize}
	\item A smaller $\ell$ makes it very efficient to estimate remainders to the energy that come from the structure of $\psi_N^{(\mathrm{J})}$. These produce the $N\ell^2$-term.
	\item A smaller $\ell$ however also makes the behaviour of every factor $f_\ell$ in $\psi_N^{(J)}$ deviate at length scales of order $\mathcal{O}(\ell)$ from the scattering solution on the whole $\Lambda$. This produces the $\ell^{-1}$ remainder.
\end{itemize}
In the rest of the section we will show how a suitable modification of $\psi_N^{(\mathrm{J})}$ allows to improve the upper bound \eqref{eq:ub_solo_jastrow}.

\subsubsection*{Perturbation of the Jastrow function and effective Hamiltonian}

Having in mind the overall goal of finding a perturbation of $\psi_N^{(\mathrm{J})}$ that reproduces the energy appearing in \eqref{eq:ub_HC}, and more technically a perturbation with a better behaviour at distances of order $\mathcal{O}(\ell)$, let us consider a
function of the form
\begin{equation} \label{eq:perturbation}
	\psi_N(x_1,\dots,x_N) = \phi_N(x_1,\dots,x_N) \psi_N^{(\mathrm{J})}(x_1,\dots,x_N).
\end{equation}
We assume that $\phi_N \in L^2_\mathrm{sym}(\Lambda^N)$ is normalized (i.e. $\|\phi_N\|_{L^2}=1$) and that it is regular enough so that $\psi_N\in \mathcal{D}[H_N^\mathrm{HC}]$. Notice that $\psi_N$ is in turn not normalized, and we will thus have to divide all expectation values by $\|\psi_N\|^2$.

The main result of this subsection is Proposition \ref{prop:derivation_H_eff} below. It shows that the optimization problem for $\psi_N$ as the trial state for $H_N^\mathrm{HC}$ can be reduced to an optimization problem for $\phi_N$ as the trial state for the effective Hamiltonian
\begin{equation} \label{eq:H_N^eff}
	H_N^\mathrm{eff}=\sum_{j=1}^N-\Delta_{x_j}+2\sum_{j<k}^N \nabla_{x_j}\left( u_\ell(x_j-x_k) \nabla_{x_j}\right) + 2\lambda_\ell \sum_{j<k}^N (\chi_\ell f_\ell^2)(x_j-x_k).
\end{equation}
Here  $u_\ell=1-f_\ell^2$, $\lambda_\ell$ was introduced in \eqref{eq:f_ell}, and $\chi_\ell$ is the characteristic function of the ball of radius $\ell$.

Considering the first term in the expansion for $\lambda_\ell$ in Lemma \ref{lemma:f_ell}, and approximating $f_\ell\simeq 1$, we see that the third sum in $H_N^\mathrm{eff}$ is an interaction term mediated by the potential
\begin{equation}
N^{-1+3\beta} V(N^\beta\cdot) \qquad \text{with}\quad N^{\beta}=\ell^{-1} \quad\text{and}\quad V=6 \mathfrak{a}\chi_1.
\end{equation}
The condition $\mathfrak{a}/N \ll \ell \ll 1$, already discussed above, imposes then $\beta \in (0,1)$, meaning that this scaling is of the type discussed in Section \ref{sect:proof_beta}. There are however corrections on top of this behaviour, namely, the contributions coming from $f_\ell-1$ and $\lambda_\ell-3\mathfrak{a}/(N\ell^3)$. These corrections will turn out to be non-negligible, as will not be the extra kinetic-type term appearing in the second sum in \eqref{eq:H_N^eff}. Nonetheless, adaptations of the methods discussed in Section \ref{sect:proof_beta} for $H_N^{(\beta)}$ will be strong guiding principles in the choice of the trial state for $H_N^{\mathrm{eff}}$.

We are then ready to state that, under suitable regularity assumptions on $\phi_N$ and up to remainders, the expectation of $H_N^\mathrm{eff}$ on $\phi_N$ captures the expectation of $H_N^\mathrm{HC}$ on $\psi_N$. We require the following bounds on $\phi_N$
\begin{equation}
\label{eq:assum_phi} 
\begin{split} 
\langle \phi_N, (-\Delta_{x_1}) \phi_N \rangle &\leq \frac{C}{N\ell}\\ \langle \phi_N, (-\Delta_{x_1})(-\Delta_{x_2}) \phi_N \rangle &\leq \frac{C}{N^2 \ell^3}  \\ 
\langle \phi_N, (-\Delta_{x_1})(- \Delta_{x_2})(- \Delta_{x_3})  \phi_N \rangle &\leq \frac{C}{N^3 \ell^4} \\
\langle \phi_N, (-\Delta_{x_1})(- \Delta_{x_2})(- \Delta_{x_3}) (-\Delta_{x_4})  \phi_N \rangle &\leq \frac{C}{N^4 \ell^6} \\
\langle \phi_N, (-\Delta_{x_1})^{3/4+\delta}  (-\Delta_{x_2})^{3/4+\delta} \dots (-\Delta_{x_n})^{3/4+ \delta} \phi_N \rangle &\leq \frac{C}{N^n \ell^{\alpha_n}} \\
\langle \phi_N, (-\Delta_{x_1}) (- \Delta_{x_2})^{3/4+\delta}  (-\Delta_{x_3})^{3/4+\delta}  \dots (-\Delta_{x_n})^{3/4+\delta} \phi_N \rangle &\leq \frac{C}{N^n \ell^{\beta_n}},
\end{split} \end{equation} 
for $\delta>0$ small enough and $n\le 6$. The sequences $\alpha_n$ and $\beta_n$ are defined by
\begin{equation}
\begin{split}
\alpha_n=\;&\Big(\frac{7}{6}+\delta\Big)n-\frac{4}{9}\Big(1-\Big(\frac{1}{2}\Big)^n\Big)\\
\beta_n=\;&\alpha_n+\frac{1}{2}-\delta.
\end{split}
\end{equation}
These are regularity bounds for $\phi_N$ used in order to estimate a number of expectation values of multiplication operators.

\begin{proposition} \label{prop:derivation_H_eff}
	Let $N^{-1+\theta} \ll \ell \ll N^{-1/2-\theta}$ for some $\theta>0$. Consider $\phi_N \in L^2_\mathrm{sym} (\Lambda^N)$ such that \eqref{eq:assum_phi} holds and $\|\phi_N\|_{L^2}=1$, and define $\psi_N=\phi_N\psi_N^{(\mathrm{J})}$ as in \eqref{eq:perturbation}. Then there exist $C,\varepsilon>0$ such that
	\begin{equation} \label{eq:derivation_H_eff}
	\begin{split}
	&\frac{\big\langle \psi_N, \sum_{j=1}^N -\Delta_{x_j} \psi_N\big\rangle}{\|\psi_N\|^2} \le \big\langle \phi_N, H_N^\mathrm{eff} \phi_N\big\rangle \\
	&\qquad-\frac{N(N-1)}{2}\big\langle\phi_N,\Big\{(H_{N-2}^\mathrm{eff}-4\pi\mathfrak{a}N)_{1,\dots,N-2}\otimes u_\ell(x_{N-1}-x_N)  \Big\}\phi_N \big\rangle\\
	&\qquad + C N^{-\varepsilon}.
	\end{split}
	\end{equation}
\end{proposition}

The proof of Proposition \ref{prop:derivation_H_eff} can be found in \cite[Section 3 and 4]{BasCenOlgPasSch-22b}, and we will discuss here the overall strategy. Before doing that, it is worth to heuristically discuss the meaning of the assumptions in \eqref{eq:assum_phi}. All the bounds are  satisfied if $\phi_N$ is the trial state $U_N^* e^{B_\eta} \Omega_+$ associated, as discussed in Section \ref{sect:proof_beta}, to the Hamiltonian of the type $H_N^{(\beta)}$ obtained by setting $f_\ell=1$ in $H_N^{\mathrm{eff}}$. In particular, the first bound in \eqref{eq:assum_phi} follows by realizing that (cft. with \eqref{eq:decay_w_beta}, recalling also that $\ell=N^{-\beta}$ in this identification)
\begin{equation*}
	\phi_N \simeq 1- \sum_{i<j}^N \frac{\mathfrak{a}}{N(|x_i-x_j|+\ell)}.
\end{equation*}
The second bound in \eqref{eq:assum_phi} is obtained by arguing that the most singular terms emerge when $\Delta_{x_2}$ hits the same variables as $\Delta_{x_1}$ does. This is however not the case for $\Delta_{x_3}$, since $\phi_N$ contains two-body correlations only. Adding the second and fourth Laplacian is therefore more costly in terms of energy than adding the first or third. The actual $\phi_N$ that we will choose at the very end will be different from the one briefly discussed here, but its overall behavior will still be governed by the same structure.

The proof of Proposition \ref{prop:derivation_H_eff} starts with the bound (see \cite[Proposition 3.1]{BasCenOlgPasSch-22b})
\begin{equation} \label{eq:intermediate_bound}
	\begin{split}
		\frac{\big\langle \psi_N, \sum_{j=1}^N -\Delta_{x_j} \psi_N\big\rangle}{\|\psi_N\|^2} \le \frac{E_\mathrm{kin}(\phi_N)+E_\mathrm{pot}(\phi_N)}{\|\psi_N\|^2}+\mathcal{E},
	\end{split}
\end{equation}
where
\begin{equation*}
	\begin{split}
		E_\mathrm{kin}(\phi_N)=\;&\sum_{j=1}^N\int |\nabla_{x_j}\phi_N(x_1,\dots,x_N)|^2 \prod_{i<j}^Nf_\ell^2(x_i-x_j)dx_1\dots dx_N\\
		E_\mathrm{pot}(\phi_N)=\,&\sum_{i<j}^N2 \lambda_\ell\int \chi_\ell(x_i-x_j)|\phi_N(x_1,\dots,x_N)|^2\prod_{r<s}^Nf_\ell^2(x_r-x_s)dx_1\dots dx_N
	\end{split}
\end{equation*}
and $\mathcal{E}$ is a remainder term which is controlled using \eqref{eq:assum_phi}. We immediately notice that, by formally replacing $f_\ell$ with one everywhere in the fraction in the right hand side of \eqref{eq:intermediate_bound}, we would obtain the expectation on $\phi_N$ of a version of $H_N^\mathrm{eff}$ in which $f_\ell$ has been set to one (and thus $u_\ell$ to zero). In order to correctly derive the full $H_N^{\mathrm{eff}}$ one has to take into account important corrections coming from the products $\prod_{i<j}^N f_\ell^2(x_i-x_j)$ appearing in the right hand side of \eqref{eq:intermediate_bound}. In particular:
\begin{itemize}
	\item Some factors of the products will produce the corrections to kinetic and potential energy in $H_N^\mathrm{eff}$. These are the factors $f_\ell^2$ that contain the variable $x_j$ in the case of the $\nabla_{x_j}$-term, and the factor $f_\ell^2(x_i-x_j)$ in the case of the $\chi_\ell(x_i-x_j)$-term.
	\item The remaining parts of the products must be dealt with by a cancellation between numerator and denominator in \eqref{eq:intermediate_bound}.
\end{itemize}

The expectation of $H_N^\mathrm{eff}$ on the right hand side of \eqref{eq:derivation_H_eff} will then come from the main part of the fraction in \eqref{eq:intermediate_bound}. This will ultimately reproduce the energy in \eqref{eq:ub_HC} after $\phi_N$ has been suitably chosen. The term in \eqref{eq:derivation_H_eff} that contains $H_{N-2}^\mathrm{eff}$ is in turn a reminder term that appears due to the need of performing effective simplifications between numerator and denominator, and whose control does not follow from the bounds in \eqref{eq:assum_phi}. Indeed, \eqref{eq:assum_phi} together with an application of the Sobolev embedding imply
\begin{equation} \label{eq:without_lower_bound}
\frac{N(N-1)}{2} 4\pi\mathfrak{a}N \big\langle \phi_N, u_\ell(x_{N-1}-x_N)\phi_N\big\rangle \le CN^2 \ell^2.
\end{equation}
Such an estimate would not allow to compute the energy up to order 1, since $\ell \gg N^{-1}$. We will however show in Proposition \ref{prop:quadratica_H_N^eff} a lower bound of the type $H_N^\mathrm{eff} \gtrsim 4\pi \mathfrak{a}N-C$. This will produce an important cancellation in the expression $H_{N-2}^\mathrm{eff}-4\pi\mathfrak{a}N$ that appears in \eqref{eq:derivation_H_eff}, ultimately allowing to gain a whole factor $N$ with respect to the estimate above.

\subsubsection*{Bounds on $H_N^\mathrm{eff}$}

We already argued that the Hamiltonian $H_N^\mathrm{eff}$ is of the type \eqref{eq:H_N_beta} with $\beta\in(0,1)$, up to some correcting terms that appear both in the kinetic and potential terms. The trial state $\phi_N$ will be chosen by analogy with what was done in Section \ref{sect:proof_beta}, with a number of modifications due to the extra terms.

The overall strategy is still to conjugate $H_N^\mathrm{eff}$ with unitary operators that extract the ground state energy up to remainders and terms that vanish on the vacuum. We first define the excitation Hamiltonian
\begin{equation}
	\mathcal{L}^{\mathrm{eff}}_N= U_N H_N^\mathrm{eff} U_N^*,
\end{equation}
where $U_N$ is the excitation map defined in \eqref{eq:def_U_N}. We then conjugate $\mathcal{L}_N^\mathrm{eff}$ with the exponential of a quadratic transformation implementing correlations through a kernel $\eta^\mathrm{eff}$, in analogy with \eqref{eq:G_N^beta}. However, the kernel $\eta^\mathrm{eff}$ has to be defined by taking care of the operators that appear in the effective Hamiltonian $H_N^\mathrm{eff}$, that is, the potential $2\lambda_\ell\chi_\ell f_\ell^2$ and the kinetic operator modified by the second sum in \eqref{eq:H_N^eff}. To this end, we consider the ground state $g_{\ell_0}$ of the (modified) scattering problem with Neumann boundary conditions
\begin{equation} \label{eq:modified_scattering}
\begin{cases}
-\nabla\left[f_\ell^2(x) \nabla g_{{\ell_0}}(x)\right]+\lambda_\ell\big(\chi_\ell f_\ell^2 g_{\ell_0}\big)(x)=\widetilde{\lambda_{\ell_0}} \big(f_\ell^2 g_{\ell_0}\big)(x)\quad& \frac{\mathfrak{a}}{N} \le |x| \le \ell_0\\
\partial_{|x|} g_{\ell_0}(x)=0\quad& |x|=\ell_0\\
g_{\ell_0}(x)=1\quad & |x|\ge \ell_0\\
g_{\ell_0}(x)=\lim_{|y|\searrow\mathfrak{a}/N}g_{\ell_0}(y)\quad & |x|<\frac{\mathfrak{a}}{N}
\end{cases}
\end{equation}
The kinetic and potential terms appearing in the left hand side of the first line above are indeed analogous to those appearing in $H_N^\mathrm{eff}$. The Neumann boundary condition is now imposed at a scale $\ell_0\in(\ell,1/2)$. A suitable choice of $\ell_0$ small enough but fixed will ensure that $g_{\ell_0}$ implements the right correlation structure up to energies of order one (not producing the correct energy appearing in \eqref{eq:ub_HC} yet).

The scattering problem \eqref{eq:modified_scattering} has in fact a remarkable structure in relation to the hard-core scattering problem \eqref{eq:f_ell}. A simple computation shows that
\begin{equation} \label{eq:identity_g}
g_{\ell_0}(x)=\frac{f_{\ell_0}(x)}{f_\ell(x)}
\end{equation}
for $\mathfrak{a}/N\le |x| \le \ell_0$, where $f_{\ell_0}$ is the solution to \eqref{eq:f_ell} with Neumann boundary condition on the ball of radius $\ell_0$ (i.e. with $\ell$ replaced by $\ell_0$). Moreover, for the eigenvalue $\widetilde{\lambda_{\ell_0}}$ appearing in \eqref{eq:modified_scattering}, we have the identity $\widetilde{\lambda_{\ell_0}}=\lambda_{\ell_0}$, where $\lambda_{\ell_0}$ is the quantity appearing in \eqref{eq:f_ell} but again with $\ell$ replaced by $\ell_0$. We then define
\begin{equation}
\eta_p^\mathrm{eff}=-N\widehat{1-g_{\ell_0}}(p),
\end{equation}
and we will implement correlations through the exponential of
\begin{equation}
B_{\eta^\mathrm{eff}}=\frac{1}{2}\sum_{p \in \Lambda_+^*} \eta_p^\mathrm{eff}(b^*_p b^*_{-p}-b_pb_{-p}).
\end{equation}
Due to the identity \eqref{eq:identity_g}, the correlations implemented by $B_{\eta^\mathrm{eff}}$ are well suited to correct the behaviour that $f_\ell$ implements in $\psi_N^{(\mathrm{J})}$ at distances of order $\ell\ll \mathcal{O}(1)$, and replace it with a similar behaviour at distances of order $\ell_0\simeq \mathcal{O}(1)$. This can be interpreted as improving the bound in \eqref{eq:ub_solo_jastrow} by replacing the $\ell^{-1}$-term with $\ell_0^{-1}$.

We thus define
\begin{equation}
\mathcal{G}_N^\mathrm{eff}=e^{-B_{\eta^\mathrm{eff}}} \mathcal{L}_N^\mathrm{eff} e^{B_{\eta^\mathrm{eff}}},
\end{equation}
and state its properties in the next Proposition. We will use the notation $\mathcal{N}_+$ already introduced in \eqref{eq:N_+_K}, as well as
\begin{equation}
\mathcal{V}_\ell=\frac{1}{2N}\sum_{\substack{p,q \in \Lambda_+^*,\; r\in \Lambda^*\\r \ne -p,-q}} \widehat V_\ell(r) a^*_{p+r}a^*_q a_p a_{q+r}
\end{equation}
with
\begin{equation}
V_\ell(x)=2N \lambda_\ell \chi_\ell(x) f^2_\ell(x).
\end{equation}
We further define the operators
\begin{equation}
\mathcal{P}^{(r)}=\sum_{p \in\Lambda_+^*}|p|^r a^*_p a_p
\end{equation}
parametrized by $r\in(1,5)$, and the quantities
\begin{equation}
\sigma_p^\mathrm{eff}=\sinh(\eta_p^\mathrm{eff}) \qquad\gamma_p^\mathrm{eff}=\cosh(\eta_p^\mathrm{eff}).
\end{equation}

\begin{proposition} \label{prop:quadratica_H_N^eff}
	Let $\ell=N^{-1+\nu}$ for some $\nu>0$ small enough, and let $\ell_0$ be small enough but fixed. Then, for any $\kappa\in(0,\nu/2)$, we have the lower bound
	\begin{equation} \label{eq:lower_bound_H_eff}
	\mathcal{G}_N^{\mathrm{eff}} \ge 4\pi \mathfrak{a}N-C(\mathcal{N}_++1)-\frac{C}{N^\kappa} \mathcal{P}^{(2+\kappa)}(\mathcal{N}_++1).
	\end{equation}
	Moreover, let us define the constant
	\begin{equation}
		\begin{split}
		C_N^{\mathrm{eff}}=\;&\frac{(N-1)}{2}\widehat V_\ell(0)+\sum_{p\in\Lambda_+^*}\bigg[p^2 (\sigma_p^\mathrm{eff})^2 -\sum_{q \in \Lambda^*}p\cdot(p+q)\widehat u_\ell(q)\eta_{p+q}^\mathrm{eff}\\
		&\qquad\qquad\qquad\qquad\quad+\widehat V_\ell(p)\Big((\sigma_p^\mathrm{eff})^2+\sigma_p^\mathrm{eff}\gamma_p^\mathrm{eff}\Big) \\
		&\qquad\qquad\qquad\qquad\quad+\frac{1}{2N}\sum_{q\in\Lambda_+^*}\widehat V_\ell (p-q)\eta_p^\mathrm{eff}\eta_q^\mathrm{eff}\bigg]
		\end{split}
	\end{equation}
	and the operator
	\begin{equation}
	\mathcal{Q}_N^\mathrm{eff}=\sum_{p\in\Lambda_+^*}\left[ F_p^\mathrm{eff}a^*_pa_p+\frac{1}{2} G_p^\mathrm{eff}(b^*_pb^*_{-p}+b_pb_{-p}) \right]
	\end{equation}
	with
	\begin{equation*}
	\begin{split}
	F_p^\mathrm{eff}=\;&p^2\left((\sigma_p^\mathrm{eff})^2+(\gamma_p^\mathrm{eff})^2\right)+(\widehat V_\ell*\widehat g_{\ell_0})(p)(\gamma_p^\mathrm{eff}+\sigma_p^\mathrm{eff})^2\\
	G_p^\mathrm{eff}=\;& 2p^2\gamma_p^\mathrm{eff} \sigma_p^\mathrm{eff}-2\sum_{q\in\Lambda^*}p\cdot (p+q)\widehat u_\ell(q)\eta^\mathrm{eff}_{p+q}+(\widehat V_\ell*\widehat g_{\ell_0})(p)(\gamma_p^\mathrm{eff}+\sigma_p^\mathrm{eff})^2.
	\end{split}
	\end{equation*}
	Then
	\begin{equation}
	\mathcal{G}_N^\mathrm{eff}=C_N^\mathrm{eff}+\mathcal{Q}_N^\mathrm{eff}+\mathcal{E}_N^\mathrm{eff},
	\end{equation}
	with
	\begin{equation}
	\pm \mathcal{E}_N^\mathrm{eff}\le \frac{C}{\sqrt{N\ell}}(\mathcal{P}^{(5/2)}+\mathcal{V}_\ell)(\mathcal{N}_++1).
	\end{equation}
\end{proposition}

The proof of Proposition \ref{prop:quadratica_H_N^eff} is discussed in \cite[Section 5]{BasCenOlgPasSch-22b}. It follows closely that of Proposition \ref{prop:quadratica_beta} for Hamiltonians of the type $H_N^{(\beta)}$, but further effort is needed in order to deal with the correction terms in $H_N^\mathrm{eff}$.

In the remaining part of this section we will discuss the implications of Proposition \ref{prop:quadratica_H_N^eff} and we will ultimately define the trial state $\phi_N$.

\subsubsection*{The choice of $\phi_N$}

Proposition \ref{prop:quadratica_H_N^eff} shows how conjugation by $e^{B_{\eta^\mathrm{eff}}}$ has the effect of extracting from $H_N^\mathrm{eff}$ the leading order energy \emph{of the original hard-core Hamiltonian}, up errors that are of order $\mathcal{O}(1)$ on states with few excitations. This is manifest in the lower bound \eqref{eq:lower_bound_H_eff}. A straightforward calculation also shows that
\begin{equation*}
	C_N^\mathrm{eff}\le4\pi \mathfrak{a}N + C_{\ell_0},
\end{equation*}
where $C_{\ell_0}>0$ depends on $\ell_0$ only.
This means in particular that choosing $\phi_N=U_N^* e^{B_{\eta^\mathrm{eff}}}\Omega_+$ would give
\begin{equation*}
	\langle \phi_N, H_N^\mathrm{eff} \phi_N \rangle = \langle\Omega_+,\mathcal{G}_N^\mathrm{eff} \Omega_+ \rangle \le 4\pi\mathfrak{a}N+C_{\ell_0}.
\end{equation*}
By separately proving that the bounds \eqref{eq:assum_phi} hold for $\phi_N=U_N^* e^{B_{\eta^\mathrm{eff}}}\Omega_+$ (see \cite[Section 7]{BasCenOlgPasSch-22b}), one is able to apply Proposition \ref{prop:derivation_H_eff} to such a $\phi_N$. The choice $\psi_N=\phi_N \psi_N^{(\mathrm{J})}$ as trial state for $H_N^\mathrm{HC}$ thus gives the upper bound
\begin{equation*}
	\begin{split}
		E_N^\mathrm{HC} \le\;& 4\pi\mathfrak{a}N + C_{\ell_0}\\
		&-\frac{N(N-1)}{2}\big\langle\phi_N,\Big\{(H_{N-2}^\mathrm{eff}-4\pi\mathfrak{a}N)_{1,\dots,N-2}\otimes u_\ell(x_{N-1}-x_N)  \Big\}\phi_N \big\rangle.
	\end{split}
\end{equation*}
The last term in the right hand side turns out to be negligible by an application of the lower bound in \eqref{eq:lower_bound_H_eff} (see \cite[Eq. (8.8) and below]{BasCenOlgPasSch-22b} for the detailed calculation). This ultimately gives
\begin{equation*}
	E_N^\mathrm{HC} \le 4\pi\mathfrak{a}N+C_{\ell_0}.
\end{equation*}

The above discussion shows that the trial state $\psi_N = \phi_N \psi_N^{(\mathrm{J})}$ with $\phi_N=U_N^* e^{B_{\eta^\mathrm{eff}}}\Omega_+$ is enough to give an upper bound which reproduces the order of the correction term in the right hand side of \eqref{eq:ub_HC}, with the non-optimal constant.
What is missing is the control of the behaviour of the trial state at distances of order $\ell_0$ (which is small enough but fixed), for which important contributions come from $\mathcal{Q}_N^\mathrm{eff}$ as well as from $\mathcal{C}_N^\mathrm{eff}$. This is done in analogy to what was discussed for $H_N^{(\beta)}$ after the first quadratic transformation, namely, we conjugate $G_N^\mathrm{eff}$ with the unitary operator $e^{B_{\tau^\mathrm{eff}}}$, where
\begin{equation}
\tau^\mathrm{eff}_p=\frac{1}{4}\log \frac{1-G_p^\mathrm{eff}/F_p^\mathrm{eff}}{1+G_p^\mathrm{eff}/F_p^\mathrm{eff}}.
\end{equation}
The complete result of the action of the operator $e^{B_{\tau^\mathrm{eff}}}$ on $\mathcal{G}_N^\mathrm{eff}$ can be found in \cite[Section 6]{BasCenOlgPasSch-22b}. It is analogous to Proposition \ref{prop:quadratica_tau}, up to modifications that take into account the differences between $F_p^\mathrm{eff},G_p^\mathrm{eff}$ and $F_p,G_p$. Here we simply state the final outcome.
\begin{proposition}\label{prop:diagonalization_H_eff}
	Let $\ell=N^{-1+\nu}$ for $\nu>0$ small enough, and let $\ell_0$ be small enough but fixed. Then
	\begin{equation}
	\begin{split}
		\langle \Omega, e^{-B_{\tau^\mathrm{eff}}} \mathcal{G}_N^\mathrm{eff} e^{B_{\tau^\mathrm{eff}}} \Omega\rangle \le\;& 4\pi \mathfrak{a}(N-1)+e_\Lambda \mathfrak{a}^2\\
		&-\frac{1}{2}\sum_{p\in \Lambda_+^*}\left[ p^2+8\pi\mathfrak{a}-\sqrt{|p|^4+ 16\pi \mathfrak{a}p^2}-\frac{(8\pi \mathfrak{a})^2}{2p^2}\right]\\
		&+CN^{-\nu/2}.
	\end{split}
	\end{equation}
\end{proposition}

This shows that choosing
\begin{equation}
	\phi_N = U_N^* e^{B_{\eta^\mathrm{eff}}} e^{B_{\tau^\mathrm{eff}}}\Omega_+,
\end{equation}
and then applying Proposition \ref{prop:derivation_H_eff}, \ref{prop:quadratica_H_N^eff}, and \ref{prop:diagonalization_H_eff} allows to conclude the proof of Theorem \ref{thm:main_GP}.

\subsection*{Acknowledgements}
The author gratefully acknowledges support from the European Research Council through the ERC-Advanced Grant CLaQS.
%
%


\begin{thebibliography}{10}
	
	\bibitem{BasCenOlgPasSch-22a}
	{\sc G.~Basti, S.~Cenatiempo, A.~Olgiati, G.~Pasqualetti, and B.~Schlein}, {\em
		Ground state energy of a bose gas in the {G}ross--{P}itaevskii regime}, J.
	Math. Phys., 63 (2022), p.~041101.
	
	\bibitem{BasCenOlgPasSch-22b}
	\leavevmode\vrule height 2pt depth -1.6pt width 23pt, {\em A second order upper
		bound for the ground state energy of a hard-sphere gas in the
		{G}ross-{P}itaevskii regime}, arXiv:2203.11917,  (2022).
	
	\bibitem{BasCenSch21}
	{\sc G.~Basti, S.~Cenatiempo, and B.~Schlein}, {\em A new second-order upper
		bound for the ground state energy of dilute bose gases}, Forum of
	Mathematics, Sigma, 9 (2021), p.~e74.
	
	\bibitem{BocBreCenSch-18}
	{\sc C.~Boccato, C.~Brennecke, S.~Cenatiempo, and B.~Schlein}, {\em Complete
		{{Bose}}\textendash{{Einstein Condensation}} in the
		{{Gross}}\textendash{{Pitaevskii Regime}}}, Commun. Math. Phys., 359 (2018),
	pp.~975--1026.
	
	\bibitem{BocBreCenSch-19}
	\leavevmode\vrule height 2pt depth -1.6pt width 23pt, {\em Bogoliubov
		{{Theory}} in the {{Gross}}-{{Pitaevskii Limit}}}, Acta Mathematica, 222
	(2019), pp.~219--335.
	
	\bibitem{BocBreCenSch-20}
	{\sc C.~Boccato, C.~Brennecke, S.~Cenatiempo, and B.~Schlein}, {\em The
		excitation spectrum of bose gases interacting through singular potentials},
	Journal of the European Mathematical Society, 22 (2020), pp.~2331--2403.
	
	\bibitem{Bog-47}
	{\sc N.~N. Bogoliubov}, {\em About the {{Theory}} of {{Superfluidity}}}, Izv.
	Akad. Nauk SSSR, 11 (1947), p.~77.
	
	\bibitem{BreSchSch-22b}
	{\sc C.~Brennecke, B.~Schlein, and S.~Schraven}, {\em Bogoliubov theory for
		trapped bosons in the gross--pitaevskii regime}, Ann. Henri Poincare, 23
	(2022), pp.~1583--1658.
	
	\bibitem{BreSchSch-22a}
	\leavevmode\vrule height 2pt depth -1.6pt width 23pt, {\em Bose–einstein
		condensation with optimal rate for trapped bosons in the gross–pitaevskii
		regime}, Mathematical Physics, Analysis and Geometry, 25 (2022).
	
	\bibitem{Din-49}
	{\sc R.~Dingle}, {\em Li. the zero-point energy of a system of particles}, The
	London, Edinburgh, and Dublin Philosophical Magazine and Journal of Science,
	40 (1949), pp.~573--578.
	
	\bibitem{Dys-57}
	{\sc F.~J. Dyson}, {\em Ground-{{State Energy}} of a {{Hard}}-{{Sphere Gas}}},
	Phys. Rev., 106 (1957), pp.~20--26.
	
	\bibitem{ErdSchYau-08}
	{\sc L.~Erd{\H{o}}s, B.~Schlein, and H.-T. Yau}, {\em Ground-state energy of a
		low-density bose gas: A second-order upper bound}, Physical Review A, 78
	(2008).
	
	\bibitem{FouGirJunMorOli-22}
	{\sc S.~Fournais, T.~Girardot, L.~Junge, L.~Morin, and M.~Olivieri}, {\em The
		ground state energy of a two-dimensional bose gas}, arXiv:2206.11100,
	(2022).
	
	\bibitem{FouSol20}
	{\sc S.~Fournais and J.~P. Solovej}, {\em {The energy of dilute Bose gases}},
	Annals of Mathematics, 192 (2020), pp.~893 -- 976.
	
	\bibitem{FouSol-21}
	{\sc S.~Fournais and J.~P. Solovej}, {\em The energy of dilute bose gases {II}:
		The general case}, arXiv:2108.12022,  (2021).
	
	\bibitem{GreSei-13}
	{\sc P.~Grech and R.~Seiringer}, {\em The {{Excitation Spectrum}} for {{Weakly
				Interacting Bosons}} in a {{Trap}}}, Commun. Math. Phys., 322 (2013),
	pp.~559--591.
	
	\bibitem{Hai-21}
	{\sc C.~Hainzl}, {\em Another proof of {BEC} in the {GP}-limit}, Journal of
	Mathematical Physics, 62 (2021), p.~051901.
	
	\bibitem{Jas-55}
	{\sc R.~Jastrow}, {\em Many-body problem with strong forces}, Phys. Rev., 98
	(1955), pp.~1479--1484.
	
	\bibitem{LeeHuaYan-57}
	{\sc T.~D. Lee, K.~Huang, and C.~N. Yang}, {\em Eigenvalues and eigenfunctions
		of a bose system of hard spheres and its low-temperature properties},
	Physical Review, 106 (1957), pp.~1135--1145.
	
	\bibitem{LewNamSerSol-15}
	{\sc M.~Lewin, P.~T. Nam, S.~Serfaty, and J.~P. Solovej}, {\em Bogoliubov
		spectrum of interacting {{Bose}} gases}, Comm. Pure Appl. Math., 68 (2015),
	pp.~413--471.
	
	\bibitem{LieSei-02}
	{\sc E.~H. Lieb and R.~Seiringer}, {\em Proof of {{Bose}}-{{Einstein
				Condensation}} for {{Dilute Trapped Gases}}}, Phys. Rev. Lett., 88 (2002),
	p.~170409.
	
	\bibitem{LieSei-06}
	\leavevmode\vrule height 2pt depth -1.6pt width 23pt, {\em Derivation of the
		{{Gross}}-{{Pitaevskii}} equation for rotating {{Bose}} gases}, Comm. Math.
	Phys., 264 (2006), pp.~505--537.
	
	\bibitem{LieSeiSolYng-05}
	{\sc E.~H. Lieb, R.~Seiringer, J.~P. Solovej, and J.~Yngvason}, {\em The
		Mathematics of the {{Bose}} Gas and Its Condensation}, vol.~34 of Oberwolfach
	Seminars, {Birkh\"auser Verlag}, 2005.
	
	\bibitem{LieSeiYng-00}
	{\sc E.~H. Lieb, R.~Seiringer, and J.~Yngvason}, {\em Bosons in a trap: {{A}}
		rigorous derivation of the {{Gross}}-{{Pitaevskii}} energy functional}, Phys.
	Rev. A, 61 (2000), p.~43602.
	
	\bibitem{LieYng-98}
	{\sc E.~H. Lieb and J.~Yngvason}, {\em Ground state energy of the low density
		bose gas}, Physical Review Letters, 80 (1998), pp.~2504--2507.
	
	\bibitem{MicNamOlg-19}
	{\sc A.~Michelangeli, P.~T. Nam, and A.~Olgiati}, {\em {Ground state energy of
			mixture of Bose gases}}, Reviews in Mathematical Physics, 31 (2019),
	p.~1950005.
	
	\bibitem{Mot-49}
	{\sc N.~Mott~F.R.S.}, {\em Iv. a contribution to the theory of liquid helium
		ii}, The London, Edinburgh, and Dublin Philosophical Magazine and Journal of
	Science, 40 (1949), pp.~61--71.
	
	\bibitem{NamNapRicTri-20}
	{\sc P.~T. Nam, M.~Napiórkowski, J.~Ricaud, and A.~Triay}, {\em Optimal rate
		of condensation for trapped bosons in the gross-pitaevskii regime},
	arXiv:2001.04364,  (2020).
	
	\bibitem{NamRouSei-16}
	{\sc P.~T. Nam, N.~Rougerie, and R.~Seiringer}, {\em Ground states of large
		bosonic systems: The {{Gross}}-{{Pitaevskii}} limit revisited}, Anal. PDE, 9
	(2016), pp.~459--485.
	
	\bibitem{NamTri-21}
	{\sc P.~T. Nam and A.~Triay}, {\em Bogoliubov excitation spectrum of trapped
		bose gases in the gross-pitaevskii regime}, arXiv:2106.11949,  (2021).
	
	\bibitem{Rou-20}
	{\sc N.~Rougerie}, {\em Scaling limits of bosonic ground states, from many-body
		to non-linear schrödinger}, EMS Surveys in Mathematical Sciences, 7 (2020),
	pp.~253--408.
	
	\bibitem{Sei-11}
	{\sc R.~Seiringer}, {\em The excitation spectrum for weakly interacting
		bosons}, Commun. Math. Phys., 306 (2011), pp.~565--578.
	
	\bibitem{YauYin-09}
	{\sc H.-T. Yau and J.~Yin}, {\em The second order upper bound for the ground
		energy of a bose gas}, J. Stat. Phys., 136 (2009), pp.~453--503.
	
\end{thebibliography}

\end{document}